\documentclass[12pt]{article}
\usepackage{epsfig}
\usepackage{amssymb}
\usepackage{amsmath}
\usepackage{amsfonts}
\usepackage{graphicx}
\usepackage{mathrsfs}
\usepackage[dvips]{color}
\usepackage{multirow}

\newcommand{\bsigma}{\boldsymbol{\sigma}}

\newcommand{\R}{\mathbb{R}}
\newcommand{\C}{\mathbb{C}}
\newcommand{\Z}{\mathbb{Z}}

\newcommand{\fa}{\mathfrak{a}}
\newcommand{\fb}{\mathfrak{b}}
\newcommand{\fc}{\mathfrak{c}}

\newcommand{\fg}{\mathfrak{g}}

\newcommand{\fs}{\mathfrak{s}}

\newcommand{\fz}{\mathfrak{z}}

\newcommand{\bg}{\mathbf{g}}

\newcommand{\bo}{{\mathbf{o}}}

\newcommand{\bzero}{\mathbf{0}}

\newcommand{\bC}{\mathbf{C}}
\newcommand{\bD}{\mathbf{D}}
\newcommand{\bE}{\mathbf{E}}
\newcommand{\bF}{\mathbf{F}}
\newcommand{\bG}{\mathbf{G}}
\newcommand{\bH}{\mathbf{H}}
\newcommand{\bI}{\mathbf{I}}

\newcommand{\bM}{\mathbf{M}}

\newcommand{\bS}{\mathbf{S}}

\newcommand{\bU}{\mathbf{U}}
\newcommand{\bV}{\mathbf{V}}

\newcommand{\cG}{\mathcal{G}}

\newcommand{\cK}{\mathcal{K}}

\newcommand{\cR}{\mathcal{R}}

\newcommand{\cV}{\mathcal{V}}

\newcommand{\cX}{\mathcal{X}}

\newcommand{\be}{\begin{equation}}
\newcommand{\ee}{\end{equation}}
\newcommand{\bea}{\begin{eqnarray}}
\newcommand{\eea}{\end{eqnarray}}
\newcommand{\nn}{\nonumber}

\newcommand{\ed}{\end{document}}

\newcommand{\bi}{\begin{itemize}}
\newcommand{\ei}{\end{itemize}}

\newcommand{\bce}{\begin{center}}
\newcommand{\ece}{\end{center}}

\newcommand{\sG}{\mathscr{G}}

\newcommand{\sT}{\mathscr{T}}

\newcommand{\IM}{{\rm Im}}

\newcommand{\bcK}{{\boldsymbol{\cK}}}

\newcommand{\bcG}{{\boldsymbol{\cG}}}

\newcommand{\bcX}{{\boldsymbol{\cX}}}

\newcommand{\bGamma}{{\boldsymbol{\Gamma}}}
\newcommand{\bDelta}{{\boldsymbol{\Delta}}}

\newcommand{\bPsi}{{\boldsymbol{\Psi}}}

\newcommand{\bfeta}{{\boldsymbol{\eta}}}

\newcommand{\for}{{\rm for}}

\begin{document}

\title{Dynamical formulation of low-energy scattering\\ in one dimension}

\author{Farhang Loran\thanks{E-mail address: loran@iut.ac.ir}~ and
Ali~Mostafazadeh\thanks{E-mail address:
amostafazadeh@ku.edu.tr}\\[6pt]
$^{*}$Department of Physics, Isfahan University of Technology, \\ Isfahan 84156-83111, Iran\\[6pt]
$^\dagger$Departments of Mathematics and Physics, Ko\c{c}
University,\\  34450 Sar{\i}yer, Istanbul, Turkey}

\date{ }
\maketitle

\begin{abstract}

The transfer matrix $\bM$ of a short-range potential may be expressed in terms of the time-evolution operator for an effective two-level quantum system with a time-dependent non-Hermitian Hamiltonian. This leads to a dynamical formulation of stationary scattering. We explore the utility of this formulation in the study of the low-energy behavior of the scattering data. In particular, for the exponentially decaying potentials, we devise a simple iterative scheme for computing terms of arbitrary order in the series expansion of $\bM$ in powers of the wavenumber. The coefficients of this series are determined in terms of a pair of solutions of the zero-energy stationary Schr\"odinger equation. We introduce a transfer matrix for the latter equation, express it in terms of the time-evolution operator for an effective two-level quantum system, and use it to obtain a perturbative series expansion for the solutions of the zero-energy stationary Schr\"odinger equation. Our approach allows for identifying the zero-energy resonances for scattering potentials in both full line and half-line with zeros of the entries of the zero-energy transfer matrix of the potential or its trivial extension to the full line. 
\vspace{6pt}




\noindent Keywords: Low-energy scattering, complex potential, zero-energy resonance, non-unitary quantum dynamics, Dyson series

\end{abstract}

\section{Introduction}

A real or complex-valued potential $v:\R\to\C$ is called a short-range potential, if for $|x|\to\infty$, $|v(x)|$ tends to zero faster than the Coulomb potential, $1/|x|$, \cite{yafaev}. A characteristic property of the short-range potentials $v(x)$ is that given any solution $\psi(x;k)$ of the corresponding Schr\"odinger equation,
	\be
	-\partial_x^2\psi(x;k)+v(x)\psi(x;k)=k^2\psi(x;k),
	\label{sch-eq}
	\ee
there are coefficient functions $A_\pm(k)$ and $B_\pm(k)$ such that
	\be
	\psi(x;k)\to A_\pm(k) e^{ikx}+B_\pm(k) e^{-ikx}~~~\for~~~x\to\pm\infty.
	\label{asymp}
	\ee
A $2\times 2$ matrix $\bM(k)$ that relates $A_\pm(k)$ and $B_\pm(k)$ according to,
	\be
	\left[\begin{array}{c}
	A_+(k)\\
	B_+(k)\end{array}\right]=\bM(k)\left[\begin{array}{c}
	A_-(k)\\
	B_-(k)\end{array}\right],
	\label{M-def}
	\ee
is called the transfer matrix of the potential $v$, \cite{tjp-2020}. It is uniquely determined by the potential, if we demand that it is independent of $A_-$ and $B_-$, \cite{epjp-2019}.
	
Among the solutions of (\ref{sch-eq}) are the so-called left/right-incident scattering solutions $\psi_{l/t}$ which fulfill,
	\bea
	\psi_l(x;k)&\to&\left\{\begin{array}{ccc}
	e^{ikx}+R^l(k)e^{-ikx}&\for& x\to-\infty,\\
	T(k)e^{ikx}&\for&x\to+\infty,
	\end{array}\right.\label{left}\\[6pt]
	\psi_r(x;k)&\to&\left\{\begin{array}{ccc}
	T(k)e^{-ikx}&\for& x\to-\infty,\\
	e^{-ikx}+R^r(k)e^{ikx}&\for&x\to+\infty.
	\end{array}\right.\label{right}
	\eea
The coefficient functions $R^{l/r}(k)$ and $T(k)$ are the left/right reflection and transmission amplitudes of the potential $v$. They are related to the entries $M_{ij}(k)$ of $\bM(k)$ according to \cite{tjp-2020}
	\begin{align}
	&R^l(k)=-\frac{M_{21}(k)}{M_{22}(k)},
	&&R^r(k)=\frac{M_{12}(k)}{M_{22}(k)},
	&&T(k)=\frac{1}{M_{22}(k)}.
	\label{RRT}
	\end{align}
Dividing $\psi_{l/r}(x;k)$ by $T(k)$, we obtain the Jost solutions $\psi_\pm(x;k)$ of (\ref{sch-eq});
	\bea
	\psi_+(x;k):=T(k)^{-1}\psi_l(x;k)&\to&\left\{\begin{array}{ccc}
	M_{22}(k)e^{ikx}-M_{21}(k)e^{-ikx}&\for& x\to-\infty,\\
	e^{ikx}&\for&x\to+\infty,
	\end{array}\right.\label{plus}\\[6pt]
	\psi_-(x;k):=T(k)^{-1}\psi_l(x;k)&\to&\left\{\begin{array}{ccc}
	e^{-ikx}&\for& x\to-\infty,\\
	M_{22}(k)e^{-ikx}+M_{12}(k)e^{ikx}&\for&x\to+\infty.
	\end{array}\right.\label{minus}
	\eea
	
Solving the scattering problem for a short-range potential $v$ means the determination of its reflection and transmission amplitudes. In view of (\ref{RRT}), this can be achieved by computing the transfer matrix of the potential \cite{jones-1941,abeles,thompson,yeh,pereyra,griffiths,sanchez}. Because $k^2$ has the interpretation of the energy in quantum mechanics, we refer to the $k\to 0$ behavior of the reflection and transmission amplitudes as the low-energy scattering properties of the potential. These are known to be sensitive to the decay rate of $|v(x)|$ as $x\to\pm\infty$, \cite{yafaev}. 

Let $L_\sigma^1(\R)$ denote the class of functions $f:\R\to\C$ such that $\int_{-\infty}^\infty (1+|x|^\sigma)|f(x)|dx<\infty$, where $\sigma$ is a nonnegative real number. Suppose that $v\in L^1_1(\R)$. Then for each $x\in\R$, the Jost solutions $\psi_\pm$ are analytic functions of $k$ in the upper complex half-plane $\C^+:=\{z\in\C\,|\, \IM(z)>0\}$, and continuous functions in $
\overline{\C^+}\setminus\{0\}:=\{z\in\C\,|\,
\IM(z)\geq 0~\&~z\neq 0~\}$. In view of (\ref{plus}) and (\ref{minus}), the same applies for (the entries of) the transfer matrix $\bM(k)$. If $v$ is an exponentially decaying function as $x\to\pm\infty$, i.e., there are positive real numbers $C,M$, and $\mu$ such that for $|x|\geq M$, $|v(x)|\leq C\,e^{-\mu|x|}$, then the Jost solutions and the transfer matrix are analytic functions of $k$ in $\{z\in\C\,|\,\IM(z)>-\mu~\&~z\neq 0\}$. This in particular implies that $0$ is an isolated singularity of $\bM(k)$, $R^{l/r}(k)$, and $T(k)$. In particular, they admit Laurent series expansions about $k=0$. The coefficients of these series determine the low-energy scattering behavior of the potential. Refs.~\cite{bolle-1985,bolle-1987} provide a comprehensive study of these coefficients and elaborate on the low-energy scattering properties of potentials belonging to $L_\sigma^1(\R)$ with $\sigma\geq 2$. Refs.~\cite{newton-1986} and \cite{aktosun-2001} explore the low-energy scattering behavior of the potentials belonging to $L_\sigma^1(\R)$ with $1<\sigma\leq 2$ and $\sigma=1$, respectively. See also \cite{yafaev,klaus-1988}.

The treatment of low-energy scattering provided in Refs.~\cite{yafaev,bolle-1985,bolle-1987,newton-1986,aktosun-2001,klaus-1988} relies on various technical results of functional analysis, which are mostly beyond the reach of typical physicists. The purpose of the present article is to offer an alternative and much more accessible approach to low-energy scattering in one dimension. It is based on a recent formulation of stationary scattering in which the transfer matrix $\bM(k)$ is expressed in terms of the time-evolution operator for a certain two-level quantum system \cite{ap-2014,pra-2014a}.  

The organization of this article is as follows. In Sec.~\ref{S2} we provide a brief review of the dynamical formulation of scattering theory in one dimension.  In Sec.~\ref{S3}, we discuss its application in the determination of the low-energy series expansion of the transfer matrix of exponentially decaying potentials.  Here we outline an iterative scheme for computing the coefficients of this series in terms of a pair of solutions of the zero-energy Schr\"odinger equation,
	\be
	-\phi''(x)+v(x)\phi(x)=0,
	\label{sch-eq-ZE}
	\ee
and provide a simple characterization of the zero-energy resonances. In Sec.~\ref{S4}, we introduce a transfer matrix $\bM_0$ for (\ref{sch-eq-ZE}), identify a corresponding two-level quantum system whose evolution operator yields $\bM_0$, and use the Dyson series expansion of the former to outline a perturbative solution of (\ref{sch-eq-ZE}). In Sec.~\ref{S5}, we discuss the extension of our results to potential scattering in the half-line, $[0,\infty)$, where the scattering problem depends on both the potential and the boundary condition at $x=0$. Here we offer a characterization of zero-energy resonances which turns out to be different for Dirichlet and non-Dirichlet boundary conditions. Finally, in Sec.~\ref{S6}, we offer a summary of our findings and present our concluding remarks.

\section{Dynamical formulation of stationary scattering in 1D}
\label{S2}

Let $\bg(x;k)$ be an invertible $2\times 2$ matrix-valued function, and 
	\begin{align}
	&\bPsi(x;k):=\bg(x;k)\left[\begin{array}{c}
	\psi(x;k)\\
	\psi'(x;k)\end{array}\right],
	&&\bV(x;k):=i\left[\begin{array}{cc}
	0 & 1\\
	v(x)-k^2 & 0\end{array}\right],
	\label{e2-1}
	\end{align}
where $v(x)$ is a real or complex short-range potential. Then, it is an elementary exercise to show that the time-independent Schr\"odinger equation (\ref{sch-eq}) is equivalent to the time-dependent Schr\"odinger equation,
	\be
	i\partial_x\bPsi(x;k)=\bH(x;k)\bPsi(x;k),
	\label{sch-eq-TD}
	\ee
for the $2\times 2$ matrix Hamiltonian,
	\be
	\bH(x;k):=\bg(x;k)\bV(x;k)\bg(x;k)^{-1}+i[\partial_x\bg(x;k)]\bg(x;k)^{-1}.
	\label{e2-2}
	\ee

If $\psi(x;k)$ is a solution of (\ref{sch-eq}), so that (\ref{asymp}) holds, we can choose $\bg(x;k)$ such that 
	\be
	\lim_{x\to\pm\infty}\bPsi(x;k)=
	\left[\begin{array}{c}
	A_\pm\\
	B_\pm\end{array}\right].
	\label{e2-3}
	\ee
The simplest choice for $\bg(x)$ that ensures (\ref{e2-3}) is
	\begin{align}
	&\bg(x;k):=\frac{1}{2k}\left[\begin{array}{cc}
	ke^{-ikx} & -i e^{-ikx}\\
	ke^{ikx} & i e^{ikx}\end{array}\right].
	\label{e2-4}
	\end{align}
Substituting this relation in (\ref{e2-1}) and (\ref{e2-2}), we have \cite{ap-2014}
	\begin{align}
	&\bPsi(x;k)=\frac{1}{2}\left[\begin{array}{c}
	e^{-ikx}[\psi(x)-ik^{-1}\psi'(x)]\\
	e^{ikx}[\psi(x)+ik^{-1}\psi'(x)]\end{array}\right],
	\label{Psi=}\\[6pt]
	&\bH(x;k)=\frac{v(x)}{2k}\left[\begin{array}{cc}
	1 & e^{-2ikx}\\
	-e^{2ikx} & -1\end{array}\right]=
	\frac{v(x)}{2k}\,e^{-ikx\bsigma_3}\bcK\,e^{ikx\bsigma_3},
	\label{H=}
	\end{align}
where 
	\be
	\bcK:=\left[\begin{array}{cc}
	1 & 1\\
	-1 & -1\end{array}\right]=\bsigma_3+i\bsigma_2,
	\label{bcK}
	\ee
and $\bsigma_j$ are the Pauli matrices;
	\begin{align}
	&\bsigma_1=\left[\begin{array}{cc}
	0 & 1\\
	1 &0\end{array}\right],
	&&\bsigma_2=\left[\begin{array}{cc}
	0 & -i\\
	i &0\end{array}\right],
	&&\bsigma_3=\left[\begin{array}{cc}
	1 & 0\\
	0 & -1\end{array}\right].
	\label{Pauli}
	\end{align}
	
The time-dependent Schr\"odinger equation (\ref{sch-eq-TD}) corresponding to the Hamiltonian (\ref{H=}) defines the dynamics of a two-level quantum system, if we let $x$ play the role of the evolution parameter, i.e., `time.' Note, however, that $\bH(x;k)$ is neither Hermitian nor diagonalizable.\footnote{If $v(x)$ is a real potential, $\bH(x;k)^\dagger=\bsigma_3\bH(x;k)\bsigma_3^{-1}$. This means that $\bH(x;k)$ is $\bsigma_3$-pseudo-Hermitian \cite{p123}. If $v(x)$ is a complex potential, $\bH(x;k)$ is $\bsigma_3$-pseudo-normal, i.e., $[\bH(x;k),\bH(x;k)^\sharp]=\bzero$, where $\bH(x;k)^\sharp:=\bsigma_3^{-1}\bH(x;k)^\dagger\bsigma_3$, \cite{ap-2014}.} Therefore, it does not generate a unitary `time-evolution.' Furthermore, because $\bcK^2=\bzero$, we have $\bH(x;k)^2=\bzero$, where $\bzero$ stands for the $2\times 2$ null matrix.

Let $\bU(x,x_0)$ be the time-evolution operator for the Hamiltonian (\ref{H=}) with $x_0\in\R$ playing the role of the initial `time,' i.e., the $2\times 2$ matrix-valued function of $x$ and $x_0$ that satisfies
	\begin{align}
	&i\partial_x\bU(x,x_0;k)=\bH(x;k)\bU(x,x_0;k),
	&&\bU(x_0,x_0;k)=\bI,
	\label{sch-eq-U}
	\end{align}
where $\bI$ labels the $2\times 2$ identity matrix. We can express the solution of (\ref{sch-eq-U}) as the Dyson series,
    \bea
    \bU(x,x_0;k)&:=&\bI+\sum_{n=1}^\infty(-i)^n\int_{x_0}^x\!\!dx_n
    \int_{x_0}^{x_n}\!\! dx_{n-1}\cdots\int_{x_0}^{x_2}\!\!dx_1
    \bH(x_n;k)\bH(x_{n-1};k)\cdots\bH(x_1;k)\nn\\
    &=:&\sT\exp\left\{-i\int_{x_0}^x\!\!d\tilde x\, \bH(\tilde x;k)\right\},
    \label{U-def}
    \eea
where $\sT$ stands for the time-ordering operation \cite{weinberg}. According to (\ref{sch-eq-U}), every solution $\bPsi(t;k)$ of the time-dependent Schr\"odinger equation (\ref{sch-eq-TD}) satisfies $\bPsi(x;k)=\bU(x,x_0;k)\bPsi(x_0;k)$. We can use this equation together with (\ref{M-def}), (\ref{e2-3}), and (\ref{U-def}) to conclude that \cite{ap-2014}
	\bea
	\bM(k)&=&\lim_{x\pm\to\pm\infty}\bU(x_+,x_-;k)=
	\sT\exp\left\{-i\int_{-\infty}^{\infty}\!\!\!dx\, \bH(x;k)\right\}\nn\\
	&=&\bI+\sum_{n=1}^\infty(-i)^n\!\!\int_{-\infty}^\infty\!\!dx_n
    	\int_{-\infty}^{x_n}\!\! dx_{n-1}\cdots\int_{-\infty}^{x_2}\!\!dx_1
    	\bH(x_n;k)\bH(x_{n-1};k)\cdots\bH(x_1;k).
	\label{M=}
	\eea
	
Because the entries of the transfer matrix determine the reflection and transmission amplitudes of the potential, Eq.~(\ref{M=}) offers an alternative route for the solution of the scattering problem for short-range potentials. Refs.~\cite{ap-2014,pra-2014a,jpa-2014a,jpa-2014b} explore various implications of this approach to potential scattering in one dimension, while Refs.~\cite{pra-2016,jpa-2020a,jpa-2020b} develop its extensions to potential scattering in two and three dimensions, electromagnetic scattering by isotropic scatterers, and potential scattering for long-range potentials. 

In the present article, we examine the utility of Eq.(\ref{M=}) in the study of the low-energy properties of the transfer matrix. This is motivated by the observation that for each $x\in\R$, the matrix Hamiltonian $\bH(x;k)$ is an analytic function of $k$ in the punctured complex $k$-plane, $\C\!\setminus\!\{0\}$, and that $0$ is a simple pole of $\bH(x;k)$.

\section{Transfer matrix at low energies}
\label{S3}

Consider a finite-range potential $v:\R\to\C$ with support $[x_-,x_+]$, so that	
	\begin{align}
	&\bU(x_-,x_-;k)=\bI,
	&&\bU(x_-,x_+;k)=\bM(k),
	\label{e3-0}
	\end{align}
and introduce 
	\begin{align}
	&\bGamma:=\left[\begin{array}{cc}
	1 & -1\\
	0 & 0\end{array}\right],
	&&\bDelta:=\left[\begin{array}{cc}
	1 & 1\\
	0 & 0\end{array}\right].
	\label{e3-1}
	\end{align}
Then it is easy to check that
	\be
	\frac{1}{2}\big(\bcK\,\bGamma+\bcK^T\bDelta\big)=\bI,
	\label{I=I}
	\ee
where $\bcK^T$ stands for the transpose of $\bcK$. 
	
If we multiply both sides of the first equation in (\ref{sch-eq-U}) from the left by $\bGamma$ and $\bDelta$ and substitute (\ref{H=}) in the resulting equations, we find
	\bea
	\partial_x\bD&=&v(x)\big[-x\,s(kx)\,\bD+x^2\,c(kx)\,\bG\big],
	\label{e3-2}\\
	\partial_x\bG&=&v(x)\big[x\,s(kx)\,\bG-\,d(kx)\,\bD\big],
	\label{e3-3}
	\eea
where $\bD$ and $\bG$ abbreviate
	\begin{align}
	&\bD(x,x_-;k):=i\bDelta\bU(x,x_-;k),
	&&\bG(x,x_-;k):=k\bGamma\bU(x,x_-;k),
	\label{D-G-def}
	\end{align}
and 
	\begin{align}
	&s(\tau):=\frac{\sin 2\tau}{2\tau}=1+\sum_{n=1}^\infty s_n \tau^{2n},
	&&s_n:=\frac{(-4)^{n}}{(2n+1)!},
	\label{S=}\\
	&c(\tau):=\frac{1-\cos 2\tau}{2\tau^2}=1+\sum_{n=1}^\infty c_n \tau^{2n},
	&&c_n:=\frac{2(-4)^{n}}{(2n+2)!},
	\label{C=}\\
	&d(\tau):=\frac{1+\cos 2\tau}{2}=1+\sum_{n=1}^\infty d_n \tau^{2n},
	&&d_n:=\frac{(-4)^{n}}{2[(2n)!]}.
	\label{D=}
	\end{align}
According to (\ref{I=I}) and (\ref{D-G-def}),
	\bea
	\bU(x,x_-;k)&=&\frac{1}{2}\left[\bcK\bGamma \bU(x,x_-;k)+
	\bcK^T\bDelta \bU(x,x_-;k)\right]\nn\\
	&=&	\frac{1}{2k}\left[\bcK\bG(x,x_-;k)-ik
	\bcK^T\bD(x,x_-;k)\right].
	\label{U=GD}
	\eea
This relation reduces the calculation of $\bU(x,x_-;k)$ to that of $\bD(x,x_-;k)$ and $\bG(x,x_-;k)$.

For each $\tilde x\in [x_-,x_+]$, we can identify $\bU(\tilde x,x_-;k)$ with the transfer matrix of the potential $\tilde v(x):=v(x)\theta(\tilde x-x)$, where $\theta(x)$ is the Heaviside step function \cite{ap-2014}. Because $\tilde v$ is a finite-range potential, either $\bU(\tilde x,x_-;k)$ is an analytic function of $k$ at $0$ or $k=0$ is a simple pole of this function.\footnote{This is generally true for exponentially decaying potentials \cite{bolle-1985}.} For $\tilde x=x\in [x_-,x_+]$, this argument implies the existence of matrix-valued functions $\bU^{(m)}(x,x_-)$ such that
	\be
	\bU(x,x_-;k)=\sum_{m=-1}^\infty \bU_m(x,x_-)\,k^m.
	\label{U-Laurent}
	\ee
Substituting this relation in (\ref{D-G-def}), we find
	\begin{align}
	&\bD(x,x_-;k)=\sum_{m=-1}^\infty \bD_m(x)\,k^m,
	&&\bG(x,x_-;k)=\sum_{m=-1}^\infty \bG_m(x)\,k^{m+1},
	\label{Laurent}
	\end{align}
where $\bD_m$ and $\bG_m$ are matrix-valued functions with vanishing second rows. The latter are also functions of $x_-$, but for brevity we do not make this explicit. In view of (\ref{U=GD}) -- (\ref{Laurent}), we can express the coefficients of the Laurent series expansion (\ref{U-Laurent}) of the evolution operator in the form,
	\be
	\bU^{(m)}(x,x_-)=\frac{1}{2}\left[\bcK\bG_m(x)-i\bcK^T\bD_m(x)\right].
	\label{bUm-def}
	\ee

Next, we note that the first relation in (\ref{e3-0}) together with (\ref{U-Laurent}) and (\ref{Laurent}) imply
	\begin{align}
	& \bD_m(x_-)=i\delta_{0m} \bDelta=
	\left\{\begin{array}{ccc}
	i\bDelta &\for & m=0,\\
	\bzero &\for&m\neq 0,\end{array}\right.
	&&
	\bG_m(x_-)=\delta_{0m}\bGamma=
	\left\{\begin{array}{ccc}
	\bGamma&\for & m=0,\\
	\bzero &\for&m\neq 0,\end{array}\right.
	\label{e3-5}
	\end{align}
where $\delta_{ij}$ stands for the Kronecker delta symbol. If we substitute (\ref{S=}) -- (\ref{D=}) and (\ref{Laurent}) in (\ref{e3-2}) and (\ref{e3-3}), and match the coefficients of the same powers of $k$ in both sides of the resulting equations, we arrive at the following equations for $\bD_m$ and $\bG_m$. 
	\begin{align}
	&\bD_{-1}(x)=\bzero
	\label{e3-Dm-eq}\\[6pt]
	&\bD'_{m+1}(x)=-x\, v(x) \big[\bD_{m+1}(x)-x\bG_{m}(x)+\bE_{m}(x)\big],
	\label{e3-D-eq}\\[6pt]
	&\bG'_{m}(x)=-v(x)\big[\bD_{m+1}(x)-x\,\bG_{m}(x)+\bF_{m}(x)\big].
	\label{e3-G-eq}
	\end{align}
Here $m\geq-1$, a prime stands for the differentiation with respect to $x$,
	\bea
	&&\bE_{-1}(x):=\bF_{-1}(x):=\bE_{0}(x):=\bF_{0}(x):=\bzero,
	\label{e3-6}\\
	&&\bE_{m}(x):=\sum_{n=1}^{\lfloor \frac{m}{2}\rfloor+1}s_nx^{2n}\bD_{m+1-2n}(x)
	-\sum_{n=1}^{\lfloor \frac{m+1}{2}\rfloor}c_nx^{2n+1}\bG_{m-2n}(x)
	~~\for~~m\geq 1,
	\label{e3-7}\\
	&&\bF_{m}(x):=\sum_{n=1}^{\lfloor \frac{m}{2}\rfloor+1}d_nx^{2n}\bD_{m+1-2n}(x)
	-\sum_{n=1}^{\lfloor \frac{m+1}{2}\rfloor}s_nx^{2n+1}\bG_{m-2n}(x)
	~~\for~~m\geq 1,
	\label{e3-8}
	\eea
and $\lfloor x\rfloor$ stands for the integer part of $x$, i.e., the greatest integer that is not greater than $x$.
	
Next, we multiply both sides of (\ref{e3-G-eq}) by $x$ and subtract the result from those of (\ref{e3-D-eq}) to obtain
	\[\bD'_{m+1}(x)-x\bG'_{m}(x)=x\,v(x)\left[\bF_{m}(x)-\bE_{m}(x)\right].\]
Integrating this equation yields
	\bea
	\bD_{m+1}(x)&=&\int_{x_-}^x\!\!d\tilde x\,\Big\{\tilde x\,v(\tilde x)\left[\bF_{m}(\tilde x)-\bE_{m}(\tilde x)\right]\Big\}+(x\partial_x-1)\bcG_m(x)+\bC_{m},
	\label{e3-9}
	\eea
where $\bcG_m(x)$ is any $2\times 2$ matrix-valued function satisfying
	\be
	\bcG_m'(x)=\bG_m(x),
	\label{bcG-def}
	\ee
and $\bC_{m}$ is a $2\times 2$ matrix which we fix by setting $x=x_-$ in (\ref{e3-9}) and making use of (\ref{e3-5}). This yields
	\be
	\bC_{m}=i\delta_{-1\,m}\,\bDelta-[x_-\bcG'_m(x_-)-\bcG_m(x_-)].
	\label{bC=}
	\ee

If we substitute (\ref{e3-9}) in (\ref{e3-G-eq}) and express the result in terms of $\bcG_m$, we discover that 
	\be
	-\bcG''_{m}(x)+v(x)\,\bcG_{m}(x)=v(x)\,\bS_{m}(x),
	\label{e3-10}
	\ee
where
	\be
	\bS_{m}(x):=\int_{x_-}^x\!\!d\tilde x
	\Big[\tilde x\, v(\tilde x)\big[\bF_{m}(\tilde x)-\bE_{m}(\tilde x)\Big]+\bF_{m}(x)+\bC_{m}.
	\label{bS-def}
	\ee
Because (\ref{e3-10}) is a second-order differential equation, we can determine $\bcG_m$ in a unique manner, if we impose a pair of initial conditions. Eqs.~(\ref{e3-5}) and (\ref{bcG-def}) provide one of these conditions, namely
	\begin{align}
	&\bcG_m'(x_-)=\delta_{0m}\,\bGamma.
	\label{e3-11a}
	\end{align}
Our analysis of low-energy scattering does not single out a second initial condition on $\bcG_m$. This reveals a certain degree of freedom in the choice of this function. We use this freedom to simplify our calculation of $\bG_m$ and $\bD_m$ by demanding that $\bcG_m$ satisfies
	\be
	x_-\bcG'_m(x_-)-\bcG_m(x_-)=i\delta_{-1\,m}\,\bDelta.
	\label{e3-11b}
	\ee
In view of (\ref{bC=}), this implies that $\bC_m=\bzero$, and  (\ref{e3-9}) and (\ref{bS-def}) become
	\bea
	\bD_{m+1}(x)&=&\int_{x_-}^x\!\!d\tilde x\,\Big\{\tilde x\,v(\tilde x)\left[\bF_{m}(\tilde x)-\bE_{m}(\tilde x)\right]\Big\}+(x\partial_x-1)\bcG_m(x),
	\label{e3-9n}\\
	\bS_{m}(x)&:=&\int_{x_-}^x\!\!d\tilde x
	\Big[\tilde x\, v(\tilde x)\big[\bF_{m}(\tilde x)-\bE_{m}(\tilde x)\Big]+\bF_{m}(x).
	\label{bS=}
	\eea
	
The initial conditions (\ref{e3-11a}) and (\ref{e3-11b}) determine a unique solution of (\ref{e3-10}). This solution takes a particularly simple form, if we express it in terms of the solutions, $\phi_1$ and $\phi_2$,  of the zero-energy Schr\"odinger equation~(\ref{sch-eq-ZE}) that are subject to the initial conditions\footnote{We do not state these conditions in the form $\phi_1(x_-)=1$, $\phi_2'(x_-)=\ell^{-1}$, and $\phi_2(x_-)-\ell^{-1}x_-=\phi_1'(x_-)=0$, because we will consider a generalization of our approach to the infinite-range exponentially decaying potentials by considering the limit $x_-\to-\infty$.}
	\be
	\begin{aligned}
	&\phi_1(x_-)-x_-\phi_1'(x_-)=1,\quad\quad&&\phi_1'(x_-)=0,\\
	&\phi_2(x_-)-x_-\phi_2'(x_-)=0,\quad\quad &&\phi_2'(x_-)=\ell^{-1}.
	\end{aligned}
	\label{IC}
	\ee
Here $\ell$ is an arbitrary positive real parameter with the dimension of length that we can identify with a relevant length scale entering the definition of the potential.\footnote{We have introduced the length scale $\ell$ to make sure $\phi_1$ and $\phi_2$ have the same physical dimension. In mathematical literature, $x$ is taken to be dimensionless, and $\ell$ is set to $1$, \cite{newton-1986}.} In terms of $\phi_1$ and $\phi_2$, the solution of the initial-value problem given by (\ref{e3-10}), (\ref{e3-11a}) and (\ref{e3-11b}) takes the form,
	\be
	\bcG_{m}(x)= -i\delta_{-1 m}\phi_1(x)\bDelta+
	\delta_{0 m}\ell\phi_2(x)\bGamma+
	\int_{x_-}^x d\tilde x\; \sG(x,\tilde x) v(\tilde x)\,\bS_{m}(\tilde x),
	\label{sol=}
	\ee
where $\sG(x,\tilde x)$ is the Green's function for the operator $-\partial_x^2+v(x)$ given by
	\be
	\sG(x,\tilde x):=
	\frac{\ell[\phi_1(x)\phi_2(\tilde x)-\phi_2(x)\phi_1(\tilde x)]}{\phi_1(x_-)}.
	\label{green}
	\ee
	
In view of (\ref{e3-6}) -- (\ref{e3-8}) and (\ref{bS=}),  $\bE_m(x)$, $\bF_m(x)$, and $\bS_m(x)$ only involve $\bG_n$ and $\bD_n$ with labels $n<m$. This together with (\ref{bcG-def}), (\ref{e3-9n}), and (\ref{sol=}) allow us to determine $\bD_m$ and $\bG_m$ iteratively. In the following we give the details of this calculation for $m\leq 1$. 

According to (\ref{e3-5}), $\bD_{-1}(x)=\bzero$. If we set $m=-1$ and $0$ in (\ref{sol=}) and (\ref{e3-9n}), and use (\ref{e3-6}), (\ref{bcG-def}), and (\ref{bS=}), we find
	\begin{align}
	&\bcG_{-1}(x)=-i\phi_1(x)\:\bDelta,
	&&\bcG_{0}(x)=\ell\phi_2(x)\bGamma,\\
	&\bG_{-1}(x)= -i\phi_1'(x)\:\bDelta,
	&&\bG_{0}(x)=\ell\phi'_2(x)\bGamma,
	\label{e3-22}\\
	&\bD_0(x)=-i[x\phi_1'(x)-\phi_1(x)]\:\bDelta,
	&&\bD_1(x)=\ell[x\phi_2'(x)-\phi_2(x)]\bGamma.
	\label{e3-23}
	\end{align}	
To determine $\bG_1$, we first use (\ref{S=}) -- (\ref{D=}), (\ref{e3-7}), (\ref{e3-8}), and (\ref{bS=}), to show that $s_1=-2/3$, $c_1=-1/3$, $d_1=-1$, and
	\begin{align}
	\bE_1(x)&
	=-\frac{i}{3}\,x^2\big[2\phi_1(x)-x\phi_1'(x)\big]\:\bDelta,\\
	\bF_1(x)&
	=-\frac{i}{3}\,x^2\big[3\phi_1(x)-x\phi_1'(x)\big]\:\bDelta,\\
	\bS_1(x)&
	=i\,\varsigma(x)\bDelta,
	\label{bS1=}
	\end{align}
where
	\be
	\varsigma(x):=-\frac{1}{3}\left\{x^2\big[3\phi_1(x)-x\phi_1'(x)\big]
	+\int_{x_-}^x d\tilde x\;\tilde x^3 v(\tilde x)\phi_1(\tilde x)\right\}.
	\label{e3-25}
	\ee	
Finally, we set $m=1$ in (\ref{sol=}) and use (\ref{bcG-def}) and (\ref{bS1=}) to show that
	\be
	\bG_1(x)=i\ell g_1(x)\:\bDelta,
	\label{e3-24}
	\ee
where 
	\be
	g_1(x):=\ell^{-1}\int_{x_-}^x\!\! d\tilde x\,\partial_x\sG(x,\tilde x)v(\tilde x)\varsigma(\tilde x).
	\label{e3-26}
	\ee
	
Having computed $\bD_m$ and $\bG_m$ for $m\leq 1$, we can use (\ref{bUm-def}),
(\ref{e3-5}), (\ref{e3-22}), (\ref{e3-23}), and (\ref{e3-24}) to show that
	\begin{align}
	&\bU^{(-1)}(x,x_-)=-\frac{i\phi_1'(x)}{2}\,\bcK,
	\label{U-1=}\\
	&\bU^{(0)}(x,x_-)=\frac{1}{2}\Big\{
	\ell\phi_2'(x)(\bI-\bsigma_1)+
	[\phi_1(x)-x\phi_1'(x)](\bI+\bsigma_1)\Big\},
	\label{U0=}\\
	&\bU^{(1)}(x,x_-)=
	\frac{i\ell}{2}\Big\{g_1(x)\bcK+
	[\phi_2(x)-x\phi_2'(x)]\bcK^T\Big\},
	\label{U+1=}
	\end{align}
where we have also benefitted from the identities
	\begin{align}
	&\bcK\,\bGamma=\bI-\bsigma_1,
	&&\bcK\,\bDelta=\bcK, 
	&&\bcK^T\bGamma=\bcK^T,
	&&\bcK^T\bDelta=\bI+\bsigma_1.
	\nn
	\end{align}
Substituting (\ref{U-1=}) -- (\ref{U+1=}) in (\ref{U-Laurent}), we have
	\bea
	\bU(x,x_-;k)&=&-\frac{i\phi_1'(x)}{2k}\,\bcK+
	\frac{1}{2}\Big\{
	\ell\phi_2'(x)(\bI-\bsigma_1)+
	[\phi_1(x)-x\phi_1'(x)](\bI+\bsigma_1)\Big\}+\nn\\
	&&\frac{ik\ell}{2}\Big\{g_1(x)\bcK+
	[\phi_2(x)-x\phi_2'(x)]\bcK^T\Big\}+O(k^2).
	\label{U-expand}
	\eea
	 
The above iterative procedure for computing the coefficients $\bU^{(m)}(x,x_-)$ of the low-energy series expansion of $\bU(x,x_-;k)$ reduces their determination to finding the solutions $\phi_1$ and $\phi_2$ of the zero-energy Schr\"odinger equation (\ref{sch-eq-ZE}) that fulfill the initial conditions (\ref{IC}). Because $v(x)$ vanishes outside the interval $[x_-,x_+]$, to each solution $\phi$ of (\ref{sch-eq-ZE}) there corresponds a pair of complex numbers, $\fa$ and $\fb$, such that $\phi(x)=\fa+\ell^{-1}\fb\,x$ for $x\leq x_-$. In particular, $\phi(x)-x\phi'(x)=\fa$ and $\phi'(x)=\ell^{-1}\fb$ for $x\leq x_-$. In light of this observation, we can establish the equivalence of (\ref{IC}) with the requirement that for all $x\in (-\infty, x_-]$,
	\begin{align}
	&\phi_1(x)-x\phi_1'(x)=1, &&\phi_1'(x)=0,
	&&\phi_2(x)-x\phi_2'(x)=0, &&\phi_2'(x)=\ell^{-1}.
	\label{IC-2x}
	\end{align}
These are in turn equivalent to the asymptotic boundary conditions,
	\be
	\begin{aligned}
	&\lim_{x\to-\infty}[\phi_1(x)-x\phi_1'(x)]=1, &&\quad\quad \lim_{x\to-\infty}\phi_1'(x)=0.\\
	&\lim_{x\to-\infty}[\phi_2(x)-x\phi_2'(x)]=0, &&\quad\quad \lim_{x\to-\infty}\phi_2'(x)=\ell^{-1}.
	\end{aligned}
	\label{IC-asym}
	\ee
	
Similarly, we can use the fact that $v(x)=0$ for $x\notin[x_-,x_+]$ to infer the existence of complex numbers $\fa_j$ and $\fb_j$, with $j\in\{1,2\}$, such that  $\phi_j(x)=\fa_j+\ell^{-1}\fb_jx$ for $x\geq x_+$. Again this implies $\phi_j(x)-x\phi_j'(x)=\fa_j$ and $\phi_j'(x)=\ell^{-1}\fb_j$ for $x\geq x_+$.  Hence,
	\begin{align}
	&\fa_j=\lim_{x\to+\infty}[\phi_j(x)-x\phi_j'(x)],
	&&\fb_j=\ell\,\lim_{x\to+\infty}\phi_j'(x).
	\label{e3-31}
	\end{align}
Because the Wronskian of $\phi_1$ and $\phi_2$, i.e., $W:=\phi_1(x)\phi_2'(x)-\phi_1'(x)\phi_2(x)$, is a constant \cite{ODE}, (\ref{IC-2x}) and (\ref{e3-31}) imply
	\be
	\fa_1\fb_2-\fa_2\fb_1=\ell
	\lim_{x\to+\infty}[\phi_1(x)\phi_2'(x)-\phi_1'(x)\phi_2(x)]=
	\ell\lim_{x\to-\infty}[\phi_1(x)\phi_2'(x)-\phi_1'(x)\phi_2(x)]=1.
	\label{ab-ba}
	\ee
It is also not difficult to see from (\ref{green}) and (\ref{e3-25}), that their right-hand sides remain the same if we replace them with their $x_-\to-\infty$ limit, so that 
	\begin{align}
	&\varsigma(x)=-\frac{1}{3}\left\{x^2\big[3\phi_1(x)-x\phi_1'(x)\big]
	+\int_{-\infty}^x d\tilde x\;\tilde x^3 v(\tilde x)\phi_1(\tilde x)\right\},
	\label{e3-25n}\\
	&\sG(x,\tilde x)=
	\frac{\ell[\phi_1(x)\phi_2(\tilde x)-\phi_2(x)\phi_1(\tilde x)]}{\lim_{x_-\to-\infty}\phi_1(x_-)}.
	\label{green-n}
	\end{align}
Furthermore, because according to (\ref{e3-26}), for all $x\in[x_-,+\infty)$, $g_1(x)=g_1(x_+)=:\fg_1$, we have
	\begin{align}
	&\fg_1=\lim_{x\to+\infty}g_1(x)=
	\ell^{-1}\int_{-\infty}^\infty\!\! d\tilde x\,\partial_x\sG(x,\tilde x)v(\tilde x)
	\varsigma(\tilde x).
	\label{e3-27}
	\end{align}
In view of these observations and Eqs.~(\ref{e3-0}), (\ref{U-expand}), and (\ref{e3-31}), we can express the low-energy expansion of the transfer matrix as follows.
	\be
	\bM(k)=-\frac{i\fb_1}{2k \ell }\,\bcK+
	\frac{1}{2}\big[
	\fb_2(\bI-\bsigma_1)+
	\fa_1(\bI+\bsigma_1)\big]+
	\frac{ik\ell}{2}\big(\fg_1\bcK+\fa_2\bcK^T\big)+O(k^2).
	\label{M-expand}
	\ee
	
Next, consider an exponentially decaying potential $v:\R\to\C$ that has an infinite range. Then $v\in L^1_1(\R)$, and we can use the results of Ref.~\cite{newton-1986} to conclude that the zero-energy Schr\"odinger equation for this potential has a pair of global solutions $\phi_1$ and $\phi_2$ satisfying the asymptotic boundary conditions (\ref{IC-asym}), and that for these solutions the limits in (\ref{e3-31}) exist, i.e., there are $\fa_j$ and $\fb_j$ fulfilling (\ref{e3-31}). This implies that $\phi_j(x)$ can grow at most linearly as $x\to\pm\infty$. Furthermore, we can view $v(x)$ as the
$x_\pm\to\pm\infty$ limit of the potentials $w_{x_-,x_+}:\R\to\C$ that are defined, for all $x_\pm\in\R$ with $x_+>x_-$, by
	\[w_{x_-,x_+}(x):=\left\{
	\begin{array}{ccc}
	v(x)&\for& x\in[x_-,x_+],\\
	0    &\for&x\notin[x_-,x_+].\end{array}\right.\]
Because $v(x)$ decays exponentially as $x\to\pm\infty$, the terms $\bD_m(x)$ and $\bG_m(x)$ in the series expansion of $\bD(x_+,x_-;k)$ and $\bG(x_+,x_-;k)$ for $w_{x_-,x_+}$ tend to finite values as $x_\pm\to\pm\infty$. This in turn implies that we can use our iterative calculation of the coefficients of the low-energy series expansion of the transfer matrix for $v(x)$ provided that for every function $f(x,x_-;k)$ appearing in the same calculation for $w_{x_-,x_+}(x)$ we replace $f(x_+,x_-;k)$ with $\lim_{x_\pm\to\pm\infty}f(x_+,x_-;k)$. In particular, (\ref{M-expand}) holds for exponentially decaying potentials provided that we define $\fa_j$, $\fb_j$, and $\fg_1$ using (\ref{e3-31}) and (\ref{e3-27}).

To obtain the low-energy expansion of the reflection and transmission amplitudes, we substitute the Laurent series for the entries of the transfer matrix in (\ref{RRT}) and express the result as power series in $k$. In view of (\ref{ab-ba}) and (\ref{M-expand}), for situations where $\fb_1\neq 0$, this gives
	\bea
	R^l(k)&=&-1-\frac{2i\fb_2\,k\ell}{\fb_1}
	+\frac{2(\fb^{2}_2+1)(k\ell)^2}{\fb_1^{2}}+O(k^3),
	\label{RL-expand}\\
	R^r(k)&=&-1-\frac{2i\fa_1\ell k}{\fb_1}
	+\frac{2(\fa^{2}_1+1)(k\ell)^2}{\fb_1^{2}}+O(k^3),
	\label{RR-expand}\\
	T(k)&=&-\frac{2ik\ell}{\fb_1}+\frac{2(\fa_1+\fb_2)(k\ell)^2}{\fb_1^2}+
	\frac{2i(\fa_1^2+\fb_2^2+\fa_1\fb_2-\fb_1\fg_1+1)(k\ell)^3}{\fb_1^3}+O(k^4).
	\label{T-expand}
	\eea 
If $\fb_1=0$, (\ref{ab-ba}) implies that $\fb_2\neq 0$ and $\fa_1=\fb_2^{-1}$. Furthermore, the term of order $(k\ell)^{-1}$ on the right-hand side of (\ref{M-expand}) vanishes. These together with (\ref{RRT}) lead to the following low-energy series expansions for the reflection and transmission amplitudes when $\fb_1=0$.
	\bea
	R^l(k)&=&\frac{\fb_2^2-1}{\fb_2^2+1}
	+\frac{2i\fb_2(\fb_2^2\fg_1-\fa_2)k\ell}{(\fb_2^{2}+1)^2}+O(k^2),
	\label{RL-expand-0}\\
	R^r(k)&=&-\frac{\fb_2^2-1}{\fb_2^2+1}
	+\frac{2i\fb_2(\fg_1-\fa_2\fb_2^2)k\ell}{(\fb_2^{2}+1)^2}+O(k^2),
	\label{Rr-expand-0}\\
	T(k)&=&\frac{2\fb_2}{\fb_2^2+1}+\frac{2i\fb_2^2(\fa_2+\fg_1)k\ell}{(\fb_2^2+1)^2}+O(k^2).
	\label{T-expand-0}
	\eea 

A couple of remarks are in order:
	\begin{enumerate}
	\item The coefficients of various powers of $k\ell$ that enter the above series expansions for the transfer matrix and the reflection and transmission amplitudes depend on the length scale $\ell$ in such a way that the terms of these series are $\ell$-independent. In view of (\ref{IC-asym}), (\ref{e3-31}), and (\ref{e3-25n}) -- (\ref{e3-27}), it is easy to see that $\fa_1$ and $\fb_2$ are independent of $\ell$;  $\fa_2$ and $\fg_1$  are proportional to $\ell^{-1}$, and $\fb_1$ is proportional to $\ell$. These imply the $\ell$-independence of the terms appearing on the right-hand sides of (\ref{M-expand}) and (\ref{RL-expand}) -- (\ref{T-expand-0}).
	\item The reflection and transmission amplitudes display different low-energy properties for $\fb_1=0$ and $\fb_2\neq 0$. In particular, unlike for $\fb_1\neq 0$, for $k\ell\to 0$, the left and right reflection amplitudes differ by a sign and the transmission amplitude tends to a nonzero value for $\fb_1=0$. We can trace this strange behavior to the fact that $\fb_1=0$ marks the presence of a zero-energy resonance \cite{yafaev,Newton-book,greenhow-1993,sassoli-1995,macri-2013}. By definition, this means that the Wronskian $W(k)$ of the Jost solutions $\psi_\pm(x;k)$ tend to zero as $k\to 0$, \cite{yafaev}. This follows from (\ref{plus}), (\ref{minus}), and (\ref{M-expand}), for they imply $W(k)=-2ik M_{22}(k)$ and $\lim_{k\to 0}[-2ik M_{22}(k)]=-\fb_1/\ell$. We state this result as the following characterization theorem for the zero-energy resonances of exponentially decaying potentials.
	\end{enumerate}
	\begin{itemize}
	\item[]{\bf Theorem~1:} Let $v:\R\to\C$ be an exponentially decaying potential and $\phi_1$ be the solution of the zero-energy Schr\"odinger equation (\ref{sch-eq-ZE}) that satisfies $\lim_{x\to-\infty}[\phi_1(x)-x\phi'(x)]=1$ and $\lim_{x\to-\infty}
\phi_1'(x)=0$. Then $v$ has a zero-energy resonance if and only if $\lim_{x\to+\infty}
\phi_1'(x)=0$.
	\end{itemize}

An exactly solvable example that we can use to check the utility of our approach to low-energy scattering is the barrier potential,
	\be
	v(x):=\left\{\begin{array}{ccc}
	\fz &\for & x\in[a,a+L],\\
	0 &\for & x\notin [a,a+L],
	\end{array}\right.
	\label{barrier}
	\ee
where $\fz$ is a possibly complex coupling constant, $a$ and $L$ are real parameters, and $L>0$. This is a finite-range potential with support $[a,a+L]$, so $x_-=a$ and $x_+=a+L$.

The standard approach for the determination of the transfer matrix of the barrier potential (\ref{barrier}) involves the solution of the corresponding time-independent Schr\"odinger equation (\ref{sch-eq}). This is quite straightforward, because this potential is piecewise constant. One obtains the general solution of (\ref{sch-eq}) in $[a,a+L]$ and matches this solution and its derivative at the boundary points, $x=a$ and $x=a+L$, to linear combinations of the plane waves, $A_\pm(k) e^{ikx}+B_\pm(k)e^{-ikx}$, which solve the Schr\"odinger equation (\ref{sch-eq}) outside $[a,a+L]$. This allows for expressing $A_+(k)$ and $B_+(k)$ in terms of $A_-(k)$ and $B_-(k)$ and determines the transfer matrix $\bM(k)$ via (\ref{M-def}). There is an alternative approach for calculating $\bM(k)$ which makes use of the dynamical formulation of scattering \cite{tjp-2020}. Both give the following expressions for the entries of the transfer matrix of (\ref{barrier}).
	\begin{align}
	M_{11}(k)&=M_{22}(-k)=
	e^{-ikL}\left[\cosh(kL\sqrt{\fz/k^2-1})-i\left(\frac{\fz}{2k^2}-1\right)
	\frac{\sinh(kL\sqrt{\fz/k^2-1})}{\sqrt{\fz/k^2-1}}\right]\nn\\
	&=-\frac{iL\fz\,\fs}{2k}+
	\fc-\frac{L^2\fz\,\fs}{2}-
	\frac{iL}{4}\Big[3\fc-(3+L^2\fz)\fs\Big]k+O(k^2),	
	\label{M1-barrier}\\
	M_{12}(k)&=M_{21}(-k)=-\frac{i\fz\,e^{-ik(2a+L)}
	\sinh(kL\sqrt{\fz/k^2-1})}{2k^2\sqrt{\fz/k^2-1}},\nn\\
	&=-\frac{iL\fz\,\fs}{2k}-\frac{L(2a+L)\fz\,\fs}{2}+
	\frac{iL}{4}\Big\{\fc+[(2a+L)^2\fz-1]\fs\Big\}k+O(k^2),
	\label{M2-barrier}
	\end{align}
where
	\begin{align}
	&\fc:=\cosh(L\sqrt\fz),
	&&\fs:=\frac{\sinh(L\sqrt\fz)}{L\sqrt\fz}.
	\label{cs-def}
	\end{align}

The barrier potential (\ref{barrier}) involves two length scales, namely $L$ and $|\fz|^{-1/2}$. According to (\ref{M1-barrier}) and (\ref{M2-barrier}), $k$ enters the expression for $M_{ij}(k)$ through $kL$ and $\fz/k^2$. This shows that a reasonable choice for the characteristic length scale $\ell$ is the largest of $L$ and $|\fz|^{-1/2}$. This is also supported by the fact that in the $k\to 0$ limit both of the physical requirements, $kL\ll 1$ and $k^2/|\fz|\ll 1$, hold.  

We can solve the zero-energy Schr\"odinger equation~(\ref{sch-eq-ZE}) for the barrier potential and identify the solutions satisfying the initial conditions (\ref{IC-asym}). For $x\in[a,a+L]$, they are given by
	\begin{align}
	&\phi_1(x)=\cosh[\sqrt\fz (x-a)],
	&&\phi_2(x)=\frac{1}{\ell}\left\{
	a\cosh[\sqrt\fz (x-a)]+\frac{\sinh[\sqrt\fz (x-a)]}{\sqrt\fz}\right\}.
	\nn
	\end{align}
Because $\phi_j(x)-x\phi_j'(x)$ and $\phi_j'(x)$ have the same values for $x\geq a+L$, we can use these relations together with (\ref{e3-31}) and (\ref{e3-25n}) -- (\ref{e3-27}) to show that 
	\begin{align}
	&\fa_1=\fc-L(a+L)\fz\,\fs,
	&&\fb_1=\ell L \fz\,\fs,
	\label{ab1-barrier}\\
	&\fa_2=-\frac{L}{\ell}\Big\{\fc+[a(a+L)\fz-1]\fs\Big\},
	&&\fb_2=\fc+a L\fz\,\fs,\\
	&\fg_1=-\frac{L}{2\ell}\Big\{\fc-[(2a^2+2aL+L^2)\fz+1]\fs\Big\}.
	\label{ab2-barrier}
	\end{align}	
Substituting these equations in (\ref{M-expand}) reproduces (\ref{M1-barrier}) and (\ref{M2-barrier}). This confirms the validity of the analysis leading to (\ref{M-expand}).

Let us also note that according to (\ref{ab1-barrier}), we have $\fb_1=0$, and the barrier potential develops a zero-energy resonance if and only if $\fs=0$. This happens whenever $\fz=-(\pi n/L)^2$ for some positive integer $n$. For these choices of the coupling constant,  (\ref{barrier}) describes a real potential well, and its transmission and reflection coefficients, $|T(k)|^2$ and $|R^{l/r}(k)|^2$, respectively tend to $1$ and $0$, for $k\to 0$, i.e., the potential is effectively reflectionless for low-energy incident waves. This behavior disappears once we perturb the coupling constant. For example, suppose that $\fz=-(\pi n/L)^2(1+\epsilon)$ for a real number $\epsilon$ such that $|\epsilon|<1$. Then for $\epsilon\neq 0$,  $|T(k)|^2\to 0$ and $|R^{l/r}(k)|^2\to 1$ as $k\to 0$, i.e., the potential begins acting as a perfect mirror for low-energy incident waves as soon as $\epsilon$ becomes nonzero, while it is reflectionless for these waves for $\epsilon=0$.

\section{Transfer matrix for the zero-energy Schr\"odinger equation}
\label{S4}

The coefficients of the low-energy expansions of the reflection and transmission amplitudes that we have derived in Sec.~\ref{S3} involve the constants $\fa_j$, $\fb_j$, and $\fg_1$ which depend on the solutions $\phi_1$ and $\phi_2$ of the zero-energy Schr\"odinger equation~(\ref{sch-eq-ZE}). In this section we pursue the approach of Sec.~\ref{S2} to introduce a transfer matrix for the zero-energy Schr\"odinger equation and express it in terms of the evolution operator for an effective two-level quantum system. This leads to a Dyson series expansion for this transfer matrix which we can use to obtain series expansions for $\fa_j$, $\fb_j$, and $\fg_1$. 

The zero-energy time-independent Schr\"odinger equation~(\ref{sch-eq-ZE}) is equivalent to the time-dependent Schr\"odinger equation (\ref{sch-eq-TD}) provided that we set $k=0$ and use $\phi(x)$ instead of $\psi(x;k)$ in (\ref{e2-1}). This  gives
	\be
	\bPsi(x;0)=\bg(x;0)\left[\begin{array}{c}
	\phi(x)\\
	\phi'(x)\end{array}\right].
	\label{e4-1}
	\ee

For  potentials $v$ belonging to $L^1_1(\R)$, the solutions of (\ref{sch-eq-ZE}) tend to polynomials of the form $\fa_\pm+\ell^{-1}\fb_\pm x$ as $x\to\pm\infty$, where $\fa_\pm$ and $\fb_\pm$ are complex numbers, and $\ell$ is an arbitrary length scale \cite{newton-1986}. This implies,
	\begin{align}
	&\lim_{x\to\pm\infty}[\phi(x)-x\phi'(x)]=\fa_\pm,
	&&\ell\!\!\lim_{x\to\pm\infty} \phi'(x)=\fb_\pm,
	\label{asym+-}
	\end{align}
which motivate the following analog of the asymptotic boundary conditions (\ref{e2-3}).	
	\be
	\lim_{x\to\pm\infty}\bPsi(x;0)=\left[\begin{array}{c}
	\fa_\pm\\
	\fb_\pm\end{array}\right].
	\label{e4-2}
	\ee
The simplest choice for $\bg(x;0)$ that is consistent with this requirement is
	\be
	\fg(x;0):=\left[\begin{array}{cc}
	1 & -x\\
	0 & \ell\end{array}\right].
	\label{g-zero=}
	\ee
In view of 	(\ref{e4-1}), it implies
	\be
	\bPsi(x;0)=\left[\begin{array}{c}
	\phi(x)-x\phi'(x)\\
	\ell\phi'(x)\end{array}\right].
	\label{Psi-x-zero=}
	\ee
Furthermore, substituting (\ref{g-zero=}) in (\ref{e2-2}) and setting $k=0$, we find
	\be
	\bH(x;0)=\bH_0(x):=-i\,v(x)\left[\begin{array}{cc}
	x & x^2/\ell\\
	-\ell & -x\end{array}\right]=-i\,x v(x)\bcX(x)\,\bcK\,\bcX(x)^{-1},
	\label{H-zero=}
	\ee
where $\bcK$ is given by (\ref{bcK}), and 
	\be
	\bcX(x):=\left[\begin{array}{cc}
	x & 0\\
	0 & \ell\end{array}\right].
	\label{bcX}
	\ee
Similarly to the matrix Hamiltonian (\ref{H=}), $\bH_0(x)$ is non-Hermitian and non-diagonalizable, and its square equals the null matrix.\footnote{If $\bfeta(x):=\bsigma_3\bcX(x)^{-2}$, the $\bfeta(x)$-pseudo-adjoint of $\bH_0(x)$, which is defined by $\bH_0(x)^\sharp:=\bfeta(x)^{-1}\bH_0(x)^\dagger
\bfeta(x)$, commutes with $\bH_0(x)$. Therefore, $\bH_0(x)$ is $\bfeta(x)$-pseudo-normal. For real-valued potentials, $\bH_0(x)^\sharp=-\bH_0(x)$, i.e., $i\bH_0(x)$ is $\bfeta(x)$-pseudo-Hermitian \cite{p123}.}

If we introduce the transfer matrix $\bM_0$ associated with the zero-energy Schr\"odinger equation~(\ref{sch-eq-ZE}) through the relation,
	\be
	\left[\begin{array}{c}
	\fa_+\\
	\fb_+\end{array}\right]=
	\bM_0\left[\begin{array}{c}
	\fa_-\\
	\fb_-\end{array}\right],
	\label{M-Zero-def}
	\ee
we can use (\ref{e4-2}) to establish,
	\be
	\bM_0=\lim_{x_\pm\to\pm\infty}\bU_0(x_+,x_-),
	\label{M-zero=}
	\ee
where $\bU_0(x,x_0)$ is the evolution operator for the matrix Hamiltonian (\ref{H-zero=}), i.e.,
	\bea
	\bU_0(x,x_0)&:=&\sT\exp\left\{-i\int_{x_0}^x\!\!d\tilde x\, \bH_0(\tilde x)\right\}\nn\\
	&:=&\bI+\sum_{n=1}^\infty(-i)^n\int_{x_0}^x\!\!dx_n
    	\int_{x_0}^{x_n}\!\! dx_{n-1}\cdots\int_{x_0}^{x_2}\!\!dx_1
    	\bH_0(x_n)\bH_0(x_{n-1})\cdots\bH_0(x_1).
    	\label{U-Zero-def}
    	\eea
	
Next, we use (\ref{bcK}) and (\ref{bcX}) to show that
	\begin{align}
	x_j\,\bcK\,\bcX(x_j)^{-1}\bcX(x_{j-1})\bcK=-(x_j-x_{j-1})\bcK.
	\end{align}
With the help of this identity and (\ref{H-zero=}), we obtain the following expression for the integrand on the right-hand side of (\ref{U-Zero-def}).
	\bea
	\bH_0(x_n)\bH_0(x_{n-1})\cdots\bH_0(x_1)&=&
	-i^n x_1 v_n(x_1,x_2,\cdots,x_n)\bcX(x_n)\bcK\,\bcX(x_1)^{-1}\nn\\
	&=&-i^n v_n(x_1,x_2,\cdots,x_n)\left[\begin{array}{cc}
	x_n & x_1x_n/\ell\\
	-\ell & -x_1\end{array}\right],
	\label{HHH}
	\eea
where $v_1(x_1):=v(x_1)$, and for $n\geq 2$,
	\bea
	v_n(x_1,x_2,\cdots,x_n)&:=&v(x_n)(x_n-x_{n-1})v(x_{n-1})
	\cdots v(x_2)(x_2-x_1)v(x_1)\nn\\
	&=&v(x_n)\prod_{j=1}^{n-1}(x_{j+1}-x_{j})v(x_{j}).
	\eea
Because $v_n(x_1,x_2,\cdots,x_n)$ is proportional to $v(x_1)v(x_2)\cdots v(x_n)$, (\ref{U-Zero-def}) gives a perturbative series expansion for the evolution operator $\bU_0(x,x_0)$. In particular, for the entries $U_{0\,ij}(x,x_0)$ of $\bU_0(x,x_0)$ we have
	\bea
	U_{0\,11}(x,x_0)&=&1-\sum_{n=1}^\infty \int_{x_0}^x\!\!dx_n
    	\int_{x_0}^{x_n}\!\! dx_{n-1}\cdots\int_{x_0}^{x_2}\!\!dx_1\:
	x_n v_n(x_1,x_2,\cdots,x_n),
	\label{U-011}\\
	U_{0\,12}(x,x_0)&=&-\frac{1}{\ell}\sum_{n=1}^\infty \int_{x_0}^x\!\!dx_n
    	\int_{x_0}^{x_n}\!\! dx_{n-1}\cdots\int_{x_0}^{x_2}\!\!dx_1\:
	x_1x_n v_n(x_1,x_2,\cdots,x_n),\\
	U_{0\,21}(x,x_0)&=&\ell\sum_{n=1}^\infty \int_{x_0}^x\!\!dx_n
    	\int_{x_0}^{x_n}\!\! dx_{n-1}\cdots\int_{x_0}^{x_2}\!\!dx_1\:
	v_n(x_1,x_2,\cdots,x_n),\\
	U_{0\,22}(x,x_0)&=&1+\sum_{n=1}^\infty \int_{x_0}^x\!\!dx_n
    	\int_{x_0}^{x_n}\!\! dx_{n-1}\cdots\int_{x_0}^{x_2}\!\!dx_1\:
	x_1 v_n(x_1,x_2,\cdots,x_n).
	\label{U-022}
	\eea
According to (\ref{M-zero=}), we can determine the entries $M_{0\,ij}$ of the zero-energy transfer matrix $\bM_0$ by substituting (\ref{U-011}) -- (\ref{U-022}) in
	\be
	M_{0\,ij}=\lim_{x_\pm\to\pm\infty} U_{0\,ij}(x_+,x_-).
	\label{Mij-zero-=}
	\ee
	
If, for $j\in\{1,2\}$, we use $\fa_{j\pm}$ and $\fb_{j\pm}$ to label the parameters $\fa_\pm$ and $\fb_\pm$ giving the asymptotic behavior of the solutions $\phi_j$ of the zero-energy Schr\"odinger equation of Sec.~\ref{S3}, we find
	\begin{align}
	&\fa_{1-}=\fb_{2-}=1, &&\fb_{1-}=\fa_{2-}=0,
	&&\fa_{j+}=\fa_j, &&\fb_{j+}=\fb_j,
	\label{aa-bb}
	\end{align}
where we have employed (\ref{IC-asym}), (\ref{e3-31}), and (\ref{asym+-}). 
In light of (\ref{M-Zero-def}) and (\ref{aa-bb}),
	\begin{align}
	&\fa_1=M_{0\,11},
	&&\fb_1=M_{0\,21},
	&&\fa_2=M_{0\,12},
	&&\fb_2=M_{0\,22}.
	\label{M-zero-ij}
	\end{align}
These equations together with (\ref{RL-expand}) -- (\ref{T-expand}), (\ref{U-011}) -- (\ref{U-022}), and (\ref{Mij-zero-=}) allow for a perturbative calculation of the low-energy expansions of the reflection and transmission amplitudes of any real or complex exponentially decaying potential up to and including quadratic terms in powers of $k\ell$ and to arbitrary order of perturbation. We can also calculate the cubic term contributing to the transmission amplitude of the potential, if we can determine the coefficient $\fg_1$ appearing in (\ref{T-expand}). In view of (\ref{e3-25n}) -- (\ref{e3-27}), this requires the calculation of $\phi_1(x)$, $\phi_2(x)$, and $\phi_1(x)-x\phi'_1(x)$ for all $x\in\R$.

We can use the identity $\bPsi(x;0)=\bU_0(x,x_0)\bPsi(x_0;0)$ together with (\ref{Psi-x-zero=}) to show that every solution $\phi$ of the zero-energy Schr\"odinger equation (\ref{sch-eq-ZE}) fulfills
	\bea
	\phi(x)&=&\left[U_{0\,11}(x,x_0)+\ell^{-1}x
	U_{0\,21}(x,x_0)\right][\phi(x_0)-x_0\phi'(x_0)]+\nn\\
	&&
	\left[\ell\,U_{0\,12}(x,x_0)+x\, U_{0\,22}(x,x_0)\right]\phi'(x_0),
	\label{phi=1}\\
	\phi(x)-x\phi'(x)&=&U_{0\,11}(x,x_0)[\phi(x_0)-x_0\phi'(x_0)]+
	\ell\,U_{0\,12}(x,x_0)\phi'(x_0).
	\label{phi=2}
	\eea
In particular,  setting $\phi=\phi_j$ in these equations, taking the limit $x_0\to-\infty$ in the resulting expressions, and making use of (\ref{IC-asym}), we have
	\bea
	&&\phi_1(x)=U_{0\,11}(x,-\infty)+\ell^{-1}x\,U_{0\,21}(x,-\infty),
	\label{phi1=}\\
	&&\phi_2(x)=U_{0\,12}(x,-\infty)+\ell^{-1}x\, U_{0\,22}(x,-\infty),
	\label{phi2=}\\
	&&\phi_1(x)-x\phi_1'(x)=U_{0\,11}(x,-\infty),
	\label{phi1=2}
	\eea
where $U_{0\,ij}(x,-\infty):=\lim_{x_0\to-\infty}U_{0\,ij}(x,x_0)$. Substituting (\ref{phi1=}) -- (\ref{phi1=2}) in (\ref{e3-25n}) -- (\ref{e3-27}) and employing (\ref{U-011}) -- (\ref{U-022}), we can compute $\fg_1$ and determine the term of order $(k\ell)^3$ in the low-energy expansion of the transmission amplitude (\ref{T-expand}).

Another consequence of (\ref{M-zero-ij}) is the identification of the condition, $\fb_1=0$, for the existence of zero-energy resonances with $M_{0\,21}=0$. This proves the following theorem.
	\begin{itemize}
	\item[]{\bf Theorem~2:} Let $v:\R\to\C$ be an exponentially decaying potential and $M_{0\,ij}$ be the entries of its zero-energy transfer matrix. Then $v$ has a zero-energy resonance if and only if $M_{0\,21}=0$.
	\end{itemize}

The delta-function potential, 
	\be
	v(x):=\fz\,\delta(x-a),
	\label{delta}
	\ee
with a possibly complex coupling constant $\fz$ and center $a\in\R$, provides a simple example which we can use to check the validity of our approach to low-energy scattering. To do this, we first note that the coupling constant $\fz$ has the dimension of inverse of length. In the absence of any other relevant length scale for this potential, we identify $\ell$ with $|\fz|^{-1}$, i.e., set $\ell:=|\fz|^{-1}$.\footnote{The parameter $a$, which also has the dimension of length is not an admissible length scale, because its value depends on the choice of the origin of our coordinates.} This  implies 
	\be
	\fz=\frac{e^{i\zeta}}{\ell},
	\label{ell-delta}
	\ee
where $\zeta$ is a real number. The treatment of the scattering problem for the delta-function potential (\ref{delta}) with a real coupling constant is a standard textbook problem. Letting the coupling constant take a complex value turns out not to cause any complications \cite{jpa-2006b}. 

The quickest way of solving the scattering problem for the delta-function potential (\ref{delta}) is to determine its transfer matrix through the use of (\ref{M=}). For this potential, the matrix Hamiltonian (\ref{H=}) takes the form
	\be
	\bH(x;k)=\frac{\fz}{2k}\,\delta(x-a)e^{-ika\bsigma_3}\bcK\, e^{ika\bsigma_3}.
	\ee
This implies $\bH(x_2;k)\bH(x_1;k)=\bzero$, the Dyson series (\ref{M=}) for the transfer matrix terminates, and we find
	\be
	\bM(x)=\bI-\frac{i\fz}{2k}e^{-ika\bsigma_3}\bcK\, e^{ika\bsigma_3}=
	\left[\begin{array}{cc}
	1-i\fz/2k & -i\fz e^{-2ika}/2k\\
	i\fz e^{2ika}/2k & 1+i\fz/2k\end{array}\right].	
	\ee
Using this equation to read off the entries of $\bM(k)$ and substituting them in (\ref{RRT}), we obtain 
	\bea
	R^l(k)&=&-\frac{\fz\,e^{2ika}}{\fz-2ik}=
	-\frac{e^{2i\hat a\, k\ell}}{1-2i e^{-i\zeta}k\ell}\nn\\
	&=&	-1-2i(\hat a+e^{-i\zeta})k\ell+2(\hat a^2+2\hat a e^{-i\zeta}+2e^{-2i\zeta})(k\ell)^2+O(k^3),	
	\label{RL-delta}\\[6pt]
	R^r(k)&=&-\frac{\fz\,e^{-2ika}}{\fz-2ik}=
	-\frac{e^{-2i\hat a\, k\ell}}{1-2i e^{-i\zeta}k\ell}\nn\\
	&=&	-1+2i(\hat a-e^{-i\zeta})k\ell+2(\hat a^2-2\hat a e^{-i\zeta}+2e^{-2i\zeta})(k\ell)^2+O(k^3),	
	\label{RR-delta}\\[6pt]
	T(k)&=&-\frac{2ik}{\fz-2ik}=-\frac{2ik\ell\,e^{-i\zeta}}{1-2ie^{-i\zeta}k\ell}\nn\\
	&=&-2i\,e^{-i\zeta}k\ell+4 e^{-2i\zeta}(k\ell)^2+8i e^{-3i\zeta}(k\ell)^3+O(k^4),
	\label{TT-delta}
	\eea
where $\hat a:=a/\ell$. 

To compare (\ref{RL-delta}) -- (\ref{TT-delta}) with the outcome of the application of (\ref{RL-expand}) -- (\ref{T-expand-0}) for the delta-function potential, we need to compute the coefficients $\fa_j$, $\fb_j$, and $\fg_1$. 

We can determine $\fa_j$ and $\fb_j$ using (\ref{M-zero-ij}) provided that we calculate the zero-energy transfer matrix $\bM_0$. First, we observe that according to (\ref{H-zero=}) and (\ref{ell-delta}), 
	\[\bH_0(x;0)=-ia \fz \delta(x-a)\bcX(a)\,\bcK\,\bcX(a)^{-1}.\]
Again because $\bH_0(x_1;0)\bH_0(x_2;0)=\bzero$, the Dyson series (\ref{U-Zero-def}) terminates, and (\ref{M-zero=}) and (\ref{M-zero-ij}) give
	\begin{align}
	&\bM_0=\bI-a\,\fz \bcX(a)\,\bcK\,\bcX(a)^{-1}=\bI-\hat a e^{i\zeta}\,\bcX(a)\bcK\,\bcX(a)^{-1}=
	\left[\begin{array}{cc}
	1- \hat a\, e^{i\zeta} & -\hat a^2 e^{i\zeta}\\
	e^{i\zeta} & 1+\hat a\, e^{i\zeta}\end{array}\right],
	\\
	&\fa_1=1- \hat a\, e^{i\zeta},\quad\quad\quad
	\fb_1=e^{i\zeta},\quad\quad\quad
	\fa_2=-\hat a^2 e^{i\zeta},\quad\quad\quad
	\fb_2=1+\hat a\, e^{i\zeta}.
	\label{aj-bj-delta}
	\end{align}
In particular, $\fb_1\neq 0$ and the low-energy series expansions for the reflection and transmission amplitudes are given by (\ref{RL-expand}) -- (\ref{T-expand}).
	
The calculation of $\fg_1$ is more involved. First, we observe that, because for the delta-function potentian (\ref{delta}), $v_n(x_1,x_2,\cdots,x_n)=0$ for $n\geq 1$, Eqs.~(\ref{U-011}) -- (\ref{U-022}) give
	\begin{align}
	&U_{0\,11}(x,-\infty)=1-\hat a\, e^{i\zeta}\theta(x-a),
	&&U_{0\,12}(x,-\infty)=-\hat a^2 e^{i\zeta}\theta(x-a),\nn\\
	&U_{0\,21}(x,-\infty)=e^{i\zeta}\theta(x-a),
	&&U_{0\,22}(x,-\infty)=1+\hat a\, e^{i\zeta}\theta(x-a).\nn
	\end{align}
Inserting these relations in (\ref{phi1=}) -- (\ref{phi1=2}), we find
	\bea
	&&\phi_1(x)=1+\left(\frac{x}{\ell}-\hat a\right)e^{i\zeta}\theta(x-a),\nn\\
	&&\phi_2(x)=\frac{x}{\ell}+\hat a
	\left(\frac{x}{\ell}-\hat a\right)e^{i\zeta}\theta(x-a),\nn\\
	&&\phi_1(x)-x\phi_1'(x)=
	1-\hat a \,e^{i\zeta}\theta(x-a).\nn
	\eea
These equations together with (\ref{e3-25n}) -- (\ref{e3-27}) imply $\fg_1=\hat a^2 e^{i\zeta}$.
Substituting this equation and (\ref{aj-bj-delta}) in (\ref{RL-expand}) -- (\ref{T-expand}) we recover (\ref{RL-delta}) -- (\ref{TT-delta}). This provides a nontrivial check on the correctness of our calculations.

\section{Low-energy scattering in the half-line}
\label{S5}
 
Consider the Schr\"odinger equation in the half-line,
	\be
	-\partial_x^2\psi(x;k)+\cV(x)\psi(x;k)=k^2\psi(x;k),~~~~~x\in\R^+,
	\label{sch-eq-half}
	\ee
where $\cV:[0,\infty)\to\C$ is a short-range potential, i.e., for $x\to+\infty$, $|\cV(x)|$ tends to zero faster than $1/x$. This implies that for every solution $\psi(x;k)$ of (\ref{sch-eq-half}), there correspond coefficient functions
$A(k)$ and $B(k)$ such that 
	\be
	\psi(x;k)\to A(k) e^{ikx}+B(k)e^{-ikx}~~~\for~~~x\to+\infty.
	\label{asym-half}
	\ee
The Schr\"odinger equation (\ref{sch-eq-half}) defines a scattering problem, if we  impose a boundary condition of the form,
	\be
	\alpha(k)\psi(0;k)+k^{-1}\beta(k)\partial_x\psi(0;k)=0,
	\label{BC}
	\ee
where $\alpha(k)$ and $\beta(k)$ are real- or complex-valued functions satisfying $|\alpha(k)|+|\beta(k)|\neq 0$. Here by the scattering problem we mean the problem of determining the reflection amplitude,
	\be
	\cR(k):=\frac{A(k)}{B(k)}.
	\label{cR}
	\ee 
	
Ref.~\cite{ap-2019} outlines a simple mapping of the scattering problem given by (\ref{sch-eq-half}) and (\ref{BC}) on the half-line to a scattering problem defined by the Schr\"odinger equation (\ref{sch-eq}) for the potential,
	\be
	v(x):=\left\{\begin{array}{ccc}
	\cV(x) & \for & x\geq 0,\\
	0 & \for & x<0,\end{array}\right.
	\label{v=half}
	\ee
in the full line. This mapping allows for relating the reflection amplitude $\cR(k)$ to the entries $M_{ij}(k)$ of the transfer matrix (alternatively the reflection and transmission amplitudes, $R^{l/r}(k)$ and $T(k)$) of the potential (\ref{v=half}) according to
	\be
	\cR(k)=\frac{M_{11}(k)-\gamma(k)M_{12}(k)}{M_{21}(k)-\gamma(k)M_{22}(k)}=
	R^r(k)-\frac{T(k)^2}{R^{l}(k)+\gamma(k)},
	\label{cR=}
	\ee
where 
	\[\gamma(k):=\frac{\alpha(k)+i\beta(k)}{\alpha(k)-i\beta(k)}.\]
For the well-known Dirichlet and Neumann boundary conditions, which correspond to setting $\beta(k)=\alpha(k)-1=0$ and $\alpha(k)=\beta(k)-1=0$, we respectively have $\gamma(k)=1$ and $\gamma(k)=-1$. 

If $\cV$ is an exponentially decaying potential, the same holds for $v$, and we can use (\ref{cR=}) to determine the low-energy expansion of the reflection amplitude $\cR(k)$ in terms of the low-energy expansion of the transfer matrix of $v$ which we have derived in Sec.~\ref{S3}. In general, this requires the knowledge of the function $\gamma(k)$. 

For situations where $\gamma$ is a constant, we can choose $\alpha$ and $\beta$ to be constant as well. In this case, we can read off the expressions for $M_{ij}(k)$ from (\ref{M-expand}) and substitute them in (\ref{cR=}) to obtain the following low-energy series expansion for $\cR(k)$. 
	\begin{align}
	&\mbox{For $\beta=0$}: 
	&&\cR(k)=\left\{\begin{array}{ccc}
	1+O(k) &\for&\fb_2=0,\\[3pt]
	-1\displaystyle- \frac{2i\fa_2\,k\ell}{\fb_2}+O(k^2) &\for&\fb_2\neq 0.\end{array}\right.
	\label{cR=2a}\\[9pt]
	&\mbox{For $\beta\neq 0$}: 
	&&\cR(k)=\left\{\begin{array}{ccc}
	-1- \displaystyle\frac{2i\fa_1\,k\ell}{\fb_1}
	+\frac{2(\fa_1^2+i\rho)(k\ell)^2}{\fb_1^2}+O(k^3)
	&\for&\fb_1\neq 0,\\[9pt]
	\displaystyle-\frac{\rho\,\fb_2^2-i}{\rho\,\fb_2^2+i}
	-\frac{2i\fb_2(\rho^2\fa_2\fb_2^2 
	+\fg_1)k\ell}{(\rho\,\fb_2^2+i)^2}+O(k^2)
	&\for&\fb_1= 0.
	\end{array}\right.
	\label{cR=2b}
	\end{align}
Here $\rho:=\alpha/\beta$, and $\fa_j$ and $\fb_j$ are the constants given by (\ref{e3-31}) provided that we identify $\phi_j$ with the solutions of the zero-energy Schr\"odinger equation (\ref{sch-eq-ZE}) for the potential (\ref{v=half}) that fulfill (\ref{IC-asym}). We can use the results of Sec.~\ref{S4} to calculate $\fa_j$, $\fb_j$, and $\fg_1$. 

According to (\ref{cR=2a}), when we impose the Dirichlelt boundary condition at $x=0$, i.e., set $\beta=0$, the reflection amplitude $\cR(k)$ displays different low-energy behavior for $\fb_2=0$ and $\fb_2\neq 0$. We can trace the root of this phenomenon to the existence of a zero-energy resonance. By definition, for the scattering problem defined by the Schr\"odinger equation (\ref{sch-eq-half}) and Dirichlet boundary condition $\psi(0;k)=0$, a zero-energy resonance arises when the Jost solution (\ref{plus}) of (\ref{sch-eq}) satisfies $\lim_{k\to 0}\psi_+(0;k)=0$, \cite{yafaev}. Because, in view of (\ref{plus}) and (\ref{M-expand}), 
	\[\lim_{k\to 0}\psi_+(0;k)=\lim_{k\to 0}[M_{22}(k)-M_{21}(k)]=\fb_2,\]
this happens provided that $\fb_2=0$. In view of (\ref{M-zero-ij}), this proves the following theorem.
	\begin{itemize}
	\item[]{\bf Theorem~3:} 
	Let $\cV:[0,\infty)\to\C$ be an exponentially decaying potential, $v:\R\to\C$ be its trivial extension to $\R$ that is given by (\ref{v=half}), $\phi_2$ be the solution of the zero-energy Schr\"odinger equation (\ref{sch-eq-ZE}) satisfying $\phi_2(0)=0$ and $\phi_2'(0)=\ell$ or some $\ell\in\R^+$, and $M_{0\,ij}$ be the entries of the zero-energy transfer matrix for $v$.  Then the following assertions are equivalent.
	\begin{enumerate}
	\item The scattering problem defined by the Schr\"odinger equation (\ref{sch-eq-half}) on the half-line and the Dirichlet boundary condition, $\psi(0;k)=0$, yields a zero-energy resonance.
	\item $\lim_{x\to+\infty}\phi_2'(x)=0$.
	\item $M_{0\,22}=0$.
	\end{enumerate}
	\end{itemize}

For the scattering problems in the half-line involving boundary conditions at $x=0$ other than Dirichlet's, we can identify the zero-energy resonances with situations where $\cR(k)$ develops a discontinuity at $k=0$. In view of (\ref{cR=2b}) this occurs when $\fb_1=0$. This establishes the following theorem.
	\begin{itemize}
	\item[]{\bf Theorem~4:} 
	Let $\cV:[0,\infty)\to\C$ be an exponentially decaying potential, $v:\R\to\C$ be its trivial extension to $\R$ that is given by (\ref{v=half}), $\phi_1$ be the solution of the zero-energy Schr\"odinger equation (\ref{sch-eq-ZE}) satisfying $\phi_1(0)=1$ and $\phi_1'(0)=0$, and $M_{0\,ij}$ be the entries of the zero-energy transfer matrix for $v$.  Then the following assertions are equivalent.
	\begin{enumerate}
	\item The scattering problem defined by the Schr\"odinger equation (\ref{sch-eq-half}) on the half-line and a boundary condition of the type (\ref{BC}) with $\beta\neq 0$ gives rise to a zero-energy resonance.
	\item $\lim_{x\to+\infty}\phi_1'(x)=0$.
	\item $M_{0\,21}=0$.
	\end{enumerate}
	\end{itemize}
		
For the delta-function potential $\cV(x)=\fz\,\delta(x-a)$ with $a>0$, the potential (\ref{v=half}) coincides with (\ref{delta}). Therefore, the coefficients $\fa_j$ and $\fb_j$ entering (\ref{cR=2a}) and (\ref{cR=2b}) are given by (\ref{aj-bj-delta}). Using these equations we obtain the following low-energy expression for the reflection amplitude of the delta-function potential in the half-line.
	\begin{align}
	&\mbox{For $\beta=0$}: 
	&&\cR(k)=\left\{\begin{array}{ccc}
	1+O(k) &\for&\fz=-1/a,\\[3pt]
	-1\displaystyle+ \frac{2i\hat a^2 k\ell}{\hat a+e^{-i\zeta}}+O(k^2) &\for&
	\fz\neq-1/a.\end{array}\right.
	\nn\\[9pt]
	&\mbox{For $\beta\neq 0$}: 
	&&\cR(k)=-1+ 2i(\hat a-e^{-i\zeta})\,k\ell+
	2\left[(\hat a_1-e^{-i\zeta})^2+
	i\rho\, e^{-2i\zeta}\right](k\ell)^2+O(k^3).\nn
	\end{align}
These formulas are in perfect agreement with the exact expression for $\cR(k)$ that we obtain by substituting (\ref{RL-delta}) -- (\ref{TT-delta}) in (\ref{cR=}), namely
	\[\cR(k)=\frac{\fz(1-\gamma\, e^{-2iak})+2ik}{
	\fz(\gamma-e^{2iak})-2i\gamma k}
	=\frac{1-\gamma\, e^{-2i\hat a\,k\ell}+2i e^{-i\zeta}k\ell}{\gamma-e^{2i\hat a\,k\ell}-2i\gamma e^{-i\zeta}k\ell}.\]
	
Unlike the delta-function potential on the full line, the delta-function potential in the half-line with $\beta=0$ (Dirichlet boundary condition) can support a zero-energy resonance. In view of Theorem~3 and (\ref{aj-bj-delta}), this happens for $\hat a e^{i\zeta}=-1$, which is equivalent to $\fz=-1/a$.  

In contrast to the delta-function potential, the barrier potential defined on the half-line by
	\be
	\cV(x):=\left\{\begin{array}{ccc}
	\fz &\for& x\in[a,a+L],\\
	0 &\for&x\notin[a,a+L],\end{array}\right.\quad\quad\quad a>0,
	\nn
	\ee
can develop zero-energy resonances for both $\beta=0$ and $\beta\neq 0$. These are respectively characterized by setting $\fb_2$ and $\fb_1$ equal to zero. In particular, for $\beta\neq 0$, the zero-energy resonances of the barrier potential on the half line coincide with the ones we found for the same potential placed on the full line. For $\beta= 0$, we can use (\ref{cs-def}) and (\ref{ab1-barrier}) to identify the condition, $\fb_2=0$, for the existence of zero-energy resonances with the solutions of the transcendental equation $\tanh(\sqrt\fz)+\sqrt{\fz}/a=0$. Because $a$ is real and positive, this equation has nonzero solutions $\fz$ that are real, negative, and infinite in number.

\section{Summary and concluding remarks}
\label{S6}

Writing the time-independent Schr\"odinger equation as the time-dependent Schr\"odinger equation for a two-level quantum system has a long history of applications in theoretical physics \cite{feshbach-1958}. For a short-range potential there is a particular way of setting up this correspondence that allows for expressing the transfer matrix of the potential in terms of the time-evolution operator for the two-level system. This leads to a dynamical formulation of stationary scattering theory in one dimensions \cite{ap-2014,pra-2014a} with a number of interesting and useful applications \cite{pra-2014a,jpa-2014a,jpa-2014b} and generalizations 
\cite{pra-2016,jpa-2020a,jpa-2020b}. In this article, we have used this formulation to develop an approach to low-energy scattering for exponentially decaying real or complex potentials in one dimension. This subject was throughly examined for real potentials in the 1980's by mathematicians \cite{bolle-1985,bolle-1987}. But the results employ certain technical tools of functional analysis which makes them extremely difficult to follow by physicists interested in performing low-energy scattering calculations. 

Our approach provides a simple iterative method of determining the coefficients of the low-energy series expansions for the transfer matrix and the reflection and transmission amplitudes of the potential. These coefficients depend on a pair of solutions of the zero-energy Schr\"odinger equation. We have introduced a transfer matrix $\bM_0$ and a corresponding quantum dynamics that yields series expansions for the coefficients of the low-energy series of the scattering data. This transfer matrix enjoys all the useful properties of the standard transfer matrix $\bM$ for the potential. In particular, it satisfies the same composition property. Furthermore, the zero-energy resonances of the potential turn out to be given by the zeros of $M_{0\,21}$ entry of $\bM_0$.

Our method also applies to potential scattering defined on the half-line $[0,\infty)$. In this case the scattering problem is specified by the choice of a potential $\cV$ in the half-line and a homogeneous boundary condition at $x=0$. We have mapped this problem to a scattering problem defined by the trivial extension of $\cV$ to the full line and used this mapping to determine the low-energy series expansion for the reflection amplitude of the corresponding scattering problem. This has led us to a useful characterization of the corresponding zero-energy resonances in terms of the zeros of the entries of the zero-energy transfer matrix $\bM_0$ for the trivial extension of $\cV$.

A class of low-energy scattering problems that we can easily apply our method to arises in the study of the scattering properties of traversable wormholes \cite{epjc-2020}. For simple models allowing for an analytic investigation, this turns out to be equivalent to the scattering problem for a scattering potential of the form $v(x)=r''(x)/r(x)$, where $r(x)$ is a given real-valued function specifying the geometry of the wormhole. For a large family of wormholes this function tends to first-degree polynomials, $a_\pm+b_\pm x$, as $x\to\pm\infty$. Because $r(x)$ solves the zero-energy Schr\"odinger equation~(\ref{sch-eq-ZE}) for $v(x)$, we can easily obtain the general solution of this equation as,
$\phi(x)=c_1 r(x)+c_2 r(x)\int_{-\infty}^x \!d\tilde x\,r(\tilde x)^{-2}$,
where $c_1$ and $c_2$ are arbitrary constants \cite{ODE}. This observation allows for the determination of the solutions $\phi_1$ and $\phi_2$ that enter the expression for the coefficients of the low-energy series expansion of the scattering data. Because only low-energy waves can pass through the wormhole's throat \cite{epjc-2020}, the ability to compute low-energy scattering properties of the wormhole is of basic importance. Our method can be directly applied to the wormhole scattering problems in which $v(x)$ is a finite-range or exponentially decaying potential. 

For situations where $v(x)$ belongs to $L^1_\sigma(\R)$ for some $\sigma\geq 2$, which encompasses the scattering of scalar waves by some well-known wormhole spacetimes, we can use our method to determine the coefficients of the leading-order and next-to-leading-order terms in the low-energy asymptotic expansion of the transmission amplitude  \cite{epjc-2020}. This follows from the fact that whenever $v$ belongs to $L^1_{2(n+\epsilon)}(\R)$ for some $n\in\Z^+$ and $\epsilon\in[0,1)$, the low-energy asymptotic expansion for the transfer matrix has the form $\sum_{m=-1}^{n-1}  \bM^{(m)}(k\ell)^m+\bo(k^{n-1+\epsilon})$, where $\bo(k^\mu)$ is a matrix-valued function of $k$ such that $\lim_{k\to 0} \bo(k^\mu)/k^\mu=\bzero$, \cite{bolle-1985}. For these potentials, our method is capable of calculating the coefficient matrices $\bM^{(m)}$.

In view of the recent progress in developing dynamical formulations of higher dimensional \cite{pra-2016}, long-range \cite{jpa-2020b},  and electromagnetic \cite{jpa-2020a} scattering, a natural direction of further research is to try to generalize the analysis of the present article to low-energy scattering problems for scalar and electromagnetic waves in higher dimensions, and for long-range potentials.

\section*{Acknowledgements}
This work has been supported by the Scientific and Technological Research Council of Turkey (T\"UB\.{I}TAK) in the framework of the project 120F061 and by Turkish Academy of Sciences (T\"UBA).

\ed

\ed
\begin{thebibliography}{99}

\bibitem{yafaev} 
D.~R.~Yafaev,  
{\em Mathematical Scattering Theory}
(AMS, Providence, 2010).

\bibitem{tjp-2020} 
A.~Mostafazadeh, 
``Transfer matrix in scattering theory: A survey of basic properties and recent developments,'' Turkish J.~Phys.\ {\bf 44}, 472-527 (2020).

\bibitem{epjp-2019} 
A.~Mostafazadeh, 
``Nonlinear scattering and its transfer matrix formulation in one dimension,'' Eur.\ Phys.\ J.~Plus {\bf 134}, 16 (2019).

\bibitem{jones-1941} 
R.~C.\ Jones,  
``A new calculus for the treatment of optical systems I. Description and discussion of the Calculus,''
J.\ Opt.\ Soc.\ Am.\ {\bf 31}, 488-493 (1941).

\bibitem{abeles} 
F.~Abel\`es, 
``Recherches sur la propagation des ondes \'electromagn\'etiques sinuso\"idales dans les milieux stratif{\i}\'es Application aux couches minces,''
Ann.\ Phys.\ (Paris) {\bf 12}, 596-640 (1950).

\bibitem{thompson} 
W.~T.~Thompson,  
``Transmission of elastic waves through a stratified solid medium,''
J.\ Appl.\ Phys.\ {\bf 21}, 89-93 (1950).

\bibitem{yeh} 
P.~Yeh, A.~Yariv, and C.-S.~Hong,
``Electromagnetic propagation in periodic stratified media. I. General theory,''
J.\ Opt.\ Soc.\ Am. {\bf 67}, 423-438 (1977).

\bibitem{pereyra} 
P.~Pereyra, 
``Resonant tunneling and band mixing in multichannel superlattices,''
Phys.\ Rev.\ Lett.\ {\bf 80}, 2677-2680 (1998).

\bibitem{griffiths} 
D.~J.~Griffiths and C.~A.~Steinke, 
``Waves in locally periodic media,''
Am.\ J.~Phys.~{\bf 69}, 137-154 (2001).

\bibitem{sanchez} 
L.\ L.~S\'anchez-Soto, J.\ J.~Monz\'ona, A.~G.~Barriuso, and 
J.~F.~Cari$\tilde{\rm n}$ena,
``The transfer matrix: A geometrical perspective,''
Phys.\ Rep.\ {\bf 513}, 191 (2012).

\bibitem{bolle-1985} 
D.~Boll\'e, F.~Gesztesy, and S.~F.~J.~Wilk, 
``A complete treatment of low-energy scattering in one dimension,''
J.~Operator Theory {\bf 13}, 3-32 (1985).

\bibitem{bolle-1987} 
D.~Boll\'e, F.~Gesztesy, and M.~Klaus, 
``Scattering theory for one-dimensional systems with $\int v(x)dx=0$,''
J.~Math.\ Anal.\ Appl.~{\bf 122}, 496-518 (1987).

\bibitem{newton-1986}
R.~G.~Newton,
``Low-energy scattering for medium-range potentials,''
J.~Math.\ Phys.\ {\bf 27}, 2720 (1986).

\bibitem{aktosun-2001}
T.~Aktosun and M.~Klaus,
``Small-energy asymptotics for the Schr\"odinger equation on the line,''
Inverse Problems {\bf 17}, 619 (2001).

\bibitem{klaus-1988} 
M.~Klaus,
``Low-energy behavior of the scattering matrix for the Schr\"odinger equation on the line,''
Inverse Problems {\bf 4}, 505-512 (1987).

\bibitem{ap-2014} 
A.~Mostafazadeh,  
``A dynamical formulation of one-dimensional scattering theory and its applications in optics,''
Ann.\ Phys.\ (NY) {\bf 341}, 77 (2014).

\bibitem{pra-2014a}  
A.~Mostafazadeh,  
``Transfer matrices as non-unitary S-matrices, multimode unidirectional invisibility, and perturbative inverse scattering,'' 
Phys.\ Rev.~A {\bf 89}, 012709 (2014). 

\bibitem{p123} 
A.~Mostafazadeh, 
``Pseudo-Hermiticity  versus  PT-Symmetry: The necessary condition for the reality of the spectrum of a non-Hermitian Hamiltonian,''
J.~Math.\ Phys.~{\bf 43}, 205-214 (2002); 
``Pseudo-Hermiticity versus PT-Symmetry II: A complete characterization of non-Hermitian Hamiltonians with a real spectrum,'' J.~Math.\ Phys. {\bf 43}, 2814-2816 (2002); 
``Pseudo-Hermiticity versus PT-Symmetry III: Equivalence of pseudo-Hermiticity and the presence of antilinear symmetries,''
J.~Math.\ Phys.\ {\bf 43}, 3944-3951 (2002).

\bibitem{weinberg} 
S.~Weinberg, 
{\em Quantum Theory of Fields, Vol.~I}
(Cambridge University Press, Cambridge, 1995).

\bibitem{jpa-2014a}
A. Mostafazadeh, 
``Adiabatic approximation, semiclassical scattering, and unidirectional invisibility,'' 
J.~Phys.\ A: Math.\ Theor.~{\bf 47}, 125301 (2014).

\bibitem{jpa-2014b}
A. Mostafazadeh, 
``Adiabatic series expansion and higher-order semiclassical approximations in scattering theory,'' 
J.~Phys.\ A: Math.\ Theor.~{\bf 47}, 345302 (2014).

\bibitem{pra-2016} 
F.~Loran and A.~Mostafazadeh, 
``Transfer matrix formulation of scattering theory in two and three dimensions,''
 Phys.\  Rev.~A {\bf 93}, 042707 (2016).

\bibitem{jpa-2020a}
F.~Loran and A.~Mostafazadeh, 
``Transfer-matrix formulation of the scattering of electromagnetic waves and broadband invisibility in three dimensions,''
J.~Phys.~A: Math.\ Theor.\ {\bf 53}, 165302 (2020).

\bibitem{jpa-2020b}
F.~Loran and A.~Mostafazadeh,
``Transfer matrix for long-range potentials,''
J.~Phys. A: Math.\ Theor.\ {\bf 53}, 395303 (2020).

\bibitem{ODE} 
W.~E.~Boyce  and R.~C.~DiPrima,
{\em Elementary Differential Equations and Boundary Value Problems} 
(Wiley, New Jersey, 2005).

\bibitem{Newton-book}
R.~G.~Newton,
{\em Scattering Theory of Waves and Particles,} 
2nd Ed., (Dover, New York, 2013).

\bibitem{greenhow-1993}
R.~C.~Greenhow,
``Is the spherical step-potential well exhibit the Ramsauer-Townsend effect?''
Am.~J.~Phys.\ {\bf 61}, 23-27 (1993).

\bibitem{sassoli-1995}
M.~Sassoli de Bianchi,
``Levinson's theorem, zero-energy resonances, and time delay in one-dimensional scattering systems,''
J.~Math.\ Phys.\ {\bf 35}, 2719-2733 (1994).

\bibitem{macri-2013}
P.~A.~Macri and R.~O.~Barrachina,
``A Jost function description of zero-energy resonance and transparency effects in elastic collisions,''
J.~Phys.~B: At.\ Mol.\ Opt.\ Phys.\ {\bf 46}, 065202 (2013).

\bibitem{jpa-2006b}
A.~Mostafazadeh, 
``Delta-function potential with a complex coupling,'' 
J.~Phys.~A: Math.\ Gen.\ {\bf 39}, 13495-13506 (2006).

\bibitem{ap-2019}
A.~Mostafazadeh, 
``Solving scattering problems in the half-line using methods developed for scattering in the full line,'' 
Ann.\ Phys.\ (NY) {\bf 411}, 167980 (2019).

\bibitem{feshbach-1958}
H.~Feshback and F.~Villars,
``Elementary relativistic wave mechanics of spin 0 and spin 1/2 particles,''
Rev.\ Mod.\ Phys.~{\bf 30}, 24-45 (1958).

\bibitem{epjc-2020}
B.~Azad, F.~Loran, and A.~Mostafazadeh,
``Transmission of low-energy scalar waves through a traversable wormhole,''
Eur.\ Phys.~J.~C {\bf 80}, 197 (2020).

\end{thebibliography}
